\documentclass[10pt,aps,prb,showpacs,superscriptaddress,twocolumn,citeautoscript]{revtex4-1}

\usepackage{graphicx}  % needed for figures
\usepackage{dcolumn}   % needed for some tables
\usepackage{bm}        % for math 
\usepackage{amssymb}  
\usepackage{amsmath}
\usepackage{hyperref,color,braket}
\usepackage{subcaption}

\newcommand{\beq}{\begin{equation}}
\newcommand{\eeq}{\end{equation}}

\newcommand{\fr}{\bm{r}}

\newcommand{\fK}{\bm{k}}
\newcommand{\fk}{\bm{k}}

\begin{document}
\title{Far-field Perfect Imaging with Time Modulated Gratings}
\author{Pawel Packo}
\affiliation{Department of Robotics and Mechatronics, AGH - University of Science and Technology, Al. A. Mickiewicza 30, 30-059 Krakow, Poland}
\author{Dani Torrent}
\email{dtorrent@uji.es}
\affiliation{GROC, UJI, Institut de Noves Tecnologies de la Imatge (INIT), Universitat Jaume I, 12071, Castell\'o, (Spain)}
\date{\today}
%%%%%%%%%%%%%%%%%%%%%%%%%%%%%%%%%%%%%%%%%%%%%%%%%%%%%%%%%%%%%%%%%%%%%%%%%%%%%%%%%%%%%%%%%%%

%%%%%%%%%%%%%%%%%%%%%%%%%%%%%%%%%%%%%%%%%%%%%%%%%%%%%%%%%%%%%%%%%%%%%%%%%%%%%%%%%%%%%%%%%%%
\begin{abstract}
We study the capabilities of time-modulated diffraction gratings as imaging devices. It is shown that a time-dependent but transversally homogeneous slab can be used to make a perfect image of an object in the far-field, since all the evanescent modes couple to propagative time-diffracted orders. It is found that, if the image to be obtained is axially symmetric, it can be recovered by measuring the time-signal at a single point, without the need of performing a spatial scan, so that time gratings can act as well as single-pixel imaging devices. In the case of having an object without axial symmetry, the time-grating can be combined with a spatial grating, and then the full image can be recovered again with a measurement at a single point. We apply the theory of compressive sensing to optimize the recovery method and numerical examples are provided. We show therefore that time-modulated gratings can be used to perfectly recover the image of an object in the far field and after measuring at a single point in the space, being therefore a promising approach to superesoloution and ultra-fast imaging.
\end{abstract}
%%%%%%%%%%%%%%%%%%%%%%%%%%%%%%%%%%%%%%%%%%%%%%%%%%%%%%%%%%%%%%%%%%%%%%%%%%%%%%%%%%%%%%%%%%%
\maketitle
%%%%%%%%%%%%%%%%%%%%%%%%%%%%%%%%%%%%%%%%%%%%%%%%%%%%%%%%%%%%%%%%%%%%%%%%%%%%%%%%%%%%%%%%%%%
The recovery of the image of an object by analyzing the shape of an undulatory field (typically acoustic or electromagnetic) that interacts with it is the most extended method for imaging. This method has an important limitation imposed by the theory of diffraction, namely, that only details of the order of the wavelength of the field can be coupled to free space. The field carries also information about details finer than the wavelength, but in the form of evanescent waves that exist only in the proximity of the object, so that these can be recovered only in the near-field\cite{goodman2005introduction}.

In order to overcome the limitations imposed by diffraction, a countless number of methods have been proposed, most of which essentially couple these evanescent waves to free propagating waves\cite{luo2003subwavelength,simonetti2006multiple,wood2006directed,kawata2008subwavelength,zhu2011holey,rogers2012super,ma2019far,orazbayev2020far}, so that the resolution of the image is finally increased. The underlying idea of these methods consists in enhancing the interaction of evanescent modes with spatially-structured objects, so that free propagating waves are excited. However, the resolution of the image is still limited by the finite size of the image-processing system, since the evanescent modes are coupled to free waves traveling along all directions in the space. Therefore, in order to fully recover the image of the object we should collect all these waves, which is obviously impossible for a physically finite system.

Due to the extraordinary capability to spatially manipulate fields, metamaterials\cite{sihvola2007metamaterials,cui2010metamaterials,zheludev2012metamaterials}, which are artificial periodic structures especially designed to control optical and acoustical fields, have been widely used to overcome the diffraction limit of conventional imaging systems\cite{casse2010super,zhu2011holey,lu2012hyperlenses,khorasaninejad2016metalenses,salami2019far}. Metamaterials have offered such a great number of extraordinary applications that the domain has evolved towards more ambitious horizons, and recently their ``time version'', that is, the temporal modulation of the materials' properties instead of their spatial modulation, is receiving increasing attention, since they present additional properties and applications not achievable by their spatial counterparts, like dynamic control of propagation\cite{pacheco2020temporal}, tunability\cite{torrent2020strong}, non-reciprocity\cite{nassar2017non,torrent2018nonreciprocal,chen2019nonreciprocal,zhu2020non} and gain\cite{torrent2018loss}. 

In this work we show that a time-modulated surface can be used to couple evanescent modes to the free space, so that this concept can be used to overcome the diffraction limit of conventional imaging systems. %Moreover, we will show that the time-version of the diffraction grating can also overcome the limitation of the finite size of the imaging system, what constitutes a remarkable advantage with respect to its spatial analogue. 
In the same way that a spatial grating excites a set of diffracted modes traveling along different directions, a time-grating excites a set of waves at different frequencies, each of which carries information about the Fourier transform of the image we want to recover. It will be shown that we can recover the image of the object either by analyzing the fundamental of these frequencies and make a spatial scan or by analyzing the temporal spectrum at a single point in the space,  showing that time-gratings are a new type of ultra-fast and single-pixel imaging devices. 

%%%%%%%%%%%%%%%%%%%%%%%%%%%%%%%%%%%%%%%%%%%%%%%%%%%%%%%%%%%%%%%%%%%%%%%%%%%%%
\section{Time Gratings as Dynamic Spatial Field Modulators }
%%%%%%%%%%%%%%%%%%%%%%%%%%%%%%%%%%%%%%%%%%%%%%%%%%%%%%%%%%%%%%%%%%%%%%%%%%%%%
Our model equation will be the scalar wave equation in free space,
\beq
\nabla^2\psi=\frac{1}{c_0^2}\partial_{tt}\psi.
\eeq
with $c_0$ being the phase speed of the wave field $\psi$.
 
 The time grating is modeled as an infinitesimally thin impedance boundary which response is time-dependent, consequently the field will be continuous at this interface while the normal component of the derivative will have a step discontinuity\cite{li2019ultrathin} given by
\beq \label{6sd}
\partial_z\psi^+-\partial_z\psi^-=\xi(t)\psi \hspace{8pt} \text{at} \hspace{8pt} z = 0,
\eeq
where $\xi$ is the time-modulated impedance of the surface and $\pm$ indicates the field $\psi$ at $z = 0^{\pm}$. We assume that the impedance is temporally modulated with repetition frequency $\nu$, so that we can expand $\xi(t)$ as a Fourier series,
\beq \label{nsy}
\xi(t)=\sum_{n=-\infty}^\infty\xi_{n}e^{-i2\pi n\nu t}.
\eeq
%with $\nu$ being the frequency used for temporal modulation of the impedance. 

Let us assume that a plane wave of unitary amplitude with transverse (grating-in-plane) wavevector $\fK$ and frequency $\omega_0$ arrives to the grating, as shown in Fig. \ref{fig832nma}. The solution for the field $\psi (z,t)$ will be given by
\beq 
\label{sna}
\psi(z,t)=e^{iq_0z}e^{-i\omega_0t}+\sum_{n}C_{n}e^{\pm iq_{n} z} e^{-i\Omega_nt},
\eeq
where $q_n=\sqrt{\Omega_n^2/c_0^2-k^2}$ (with $k = |\fk|$) and $\Omega_n=\omega_0+2n\pi\nu$. The $\pm$ in \eqref{sna} refers to either $z > 0$ or $z < 0$. The above solution already satisfies the field continuity condition, while the discontinuity of the derivative condition yields
\beq
2iq_{n}C_{n}=\sum_{m}\xi_{n-m}(\delta_{m0}+C_{m}),
\eeq
so that we can solve for the $C_{m}$ coefficients from the system of equations
\beq
\sum_{m}(2iq_{n}\delta_{nm}-\xi_{n-m})C_{m}=\xi_n .
\eeq
The field behind, i.e. $z > 0$, the impedance plane will be composed of a set of diffracted waves with amplitudes
\beq 
\label{asdfc}
T_{n}=\delta_{n0}+C_{n}.
\eeq
Therefore, the periodic temporal modulation of the impedance of the surface excites a set of diffracted waves in a similar way as the periodic spatial modulation, and it behaves as a diffraction grating. It is clearly an ``active'' grating in the sense that energy is not conserved in this process, since the time-dependence in the parameters of the grating has to be induced externally, and some energy has to be supplied to the system\cite{torrent2018loss}.

%{\color{red} Plot of the schematics}
%
%\begin{figure}[h!]
%	\includegraphics[width=0.5\linewidth]{figures/scheme}
%	\caption{Time-space modulated grating of impedance $\xi (\fr, t)$ spanning the $x-y$ plane, located at $z = 0$. The wave of temporal frequency $\omega_0$ and the in-plane wavevector of $\bf{k}$ impinges the grating from $z^-$ direction.}
%	\label{fig1}
%\end{figure}
%
%{\color{red} Plot of the coefficients for k=0 sinusoidal spatial modulation and sinusoidal time?Or random grating?}

The above equations have been derived assuming that the incident field is a monochromatic plane wave of unitary amplitude. In a more generic case, the incident field will be made up by a linear combination of plane waves of the form,
\beq
\psi_0(\fr,t)=\iint A_0(\fK)e^{i\fK\cdot\fr}e^{iq_0z}e^{-i\omega_0t}d^2\fK,
\eeq
where $\fr$ is a 2-D vector defined in the $x-y$ plane. The transmitted field will be
\beq
\psi_T(\fr,t)=\sum_{n}\iint A_0(\fK)T_{n}(k,\omega_0)e^{i\fK\cdot\fr}e^{-i\Omega_nt}d^2\fK,
\eeq
which can be expressed as
\beq
\psi_T(\fr,t)=\sum_ne^{-i\Omega_nt}\iint A_0(\fr')T_n(\fr,\fr')d^2\fr',
\eeq
with
\beq
T_n(\fr,\fr')=\iint T_{n}(k,\omega_0)e^{i\fK\cdot(\fr-\fr')}d^2\fK.
\eeq
It can be therefore seen that the role of the grating is to generate an infinite set of harmonics whose field distributions are different, since the transmittance of the grating is different for each harmonic. Interestingly, this is equivalent to say that the grating behaves as a spatial light modulator if the field comprises of optical waves or spatial field modulator in the most general case. Spatial light modulators have been widely used in optics for a countless number of applications, however they need to be spatially modified. In this case, we modulate a signal by a grating and send the harmonics at once with all the possible patterns. We will see later how this effect can be exploited to recover the image of an object after measuring at a single point.

Although the above problem is scalar, the fundamental principles described by it will apply as well to vectorial waves, like elastic or electromagnetic. The main difference will be the calculation of the transmission and reflection coefficients by the grating, but it is obvious that the principles of diffraction and imaging, as explained before, will be identical. Consequently, we limit the study to the scalar wave equation but bearing in mind that the results presented here will be very similar for vectorial waves.

%%%%%%%%%%%%%%%%%%%%%%%%%%%%%%%%%%%%%%%%%%%%%%%%%%%%%%%%%%%%%%%%%%%%%%%%%%%%%
%%%%%%%%%%%%%%%%%%%%%%%%%%%%%%%%%%%%%%%%%%%%%%%%%%%%%%%%%%%%%%%%%%%%%%%%%%%%%
\section{Far-field Imaging with a time grating}

Let us consider now that we want to recover the image of an object illuminated by a plane wave of frequency $\omega_0$. We define the image as a 2-D screen such that just after the screen the field distribution is $\psi_0(\fr,z=0)$, so that a set of plane waves which amplitudes can be obtained from the Fourier transform $A_0(\fk)$ of $\psi_0(\fr,z=0)$ is propagating along the $z$ axis. However, for those wavenumbers such that $k>k_0$ the propagation along $z$ is evanescent, since $q_0$ is a complex number. Consequently, the information carried by the amplitude $A_0(\fK)$ for these wavenumbers ($k>k_0$) exists only in the near field and will be lost upon propagation, therefore we will not be able to fully recover the image at a certain (rather small) distance from the source. 

We can avoid this loss of information by putting a time grating just after the object at $z=0$. The grating's response $T_n(k)$ depends as well on $\omega_0$, omitted for simplicity, but it depends only on the wavenumber modulus $k$, since at the moment we assume the grating is homogeneous (i.e. its impedance does not depend on position) and it cannot distinguish an image from its rotated version. However, the response of the grating implies that for all $\fk$ there will be always a diffraction order $n$ such that $q_n$ will be real and, consequently, the information carried by $A_0(\fk)$ will not be lost at any distance from the source. Therefore, higher harmonics are responsible for passing higher $k$'s beyond the imaging grating as propagating modes, up to physical constraints related to attenuation of that wave components in a practical realization of the system.

\begin{figure}[ht]
	\begin{subfigure}{.5\textwidth}
		\centering
		% include first image
		\includegraphics[width=.98\linewidth]{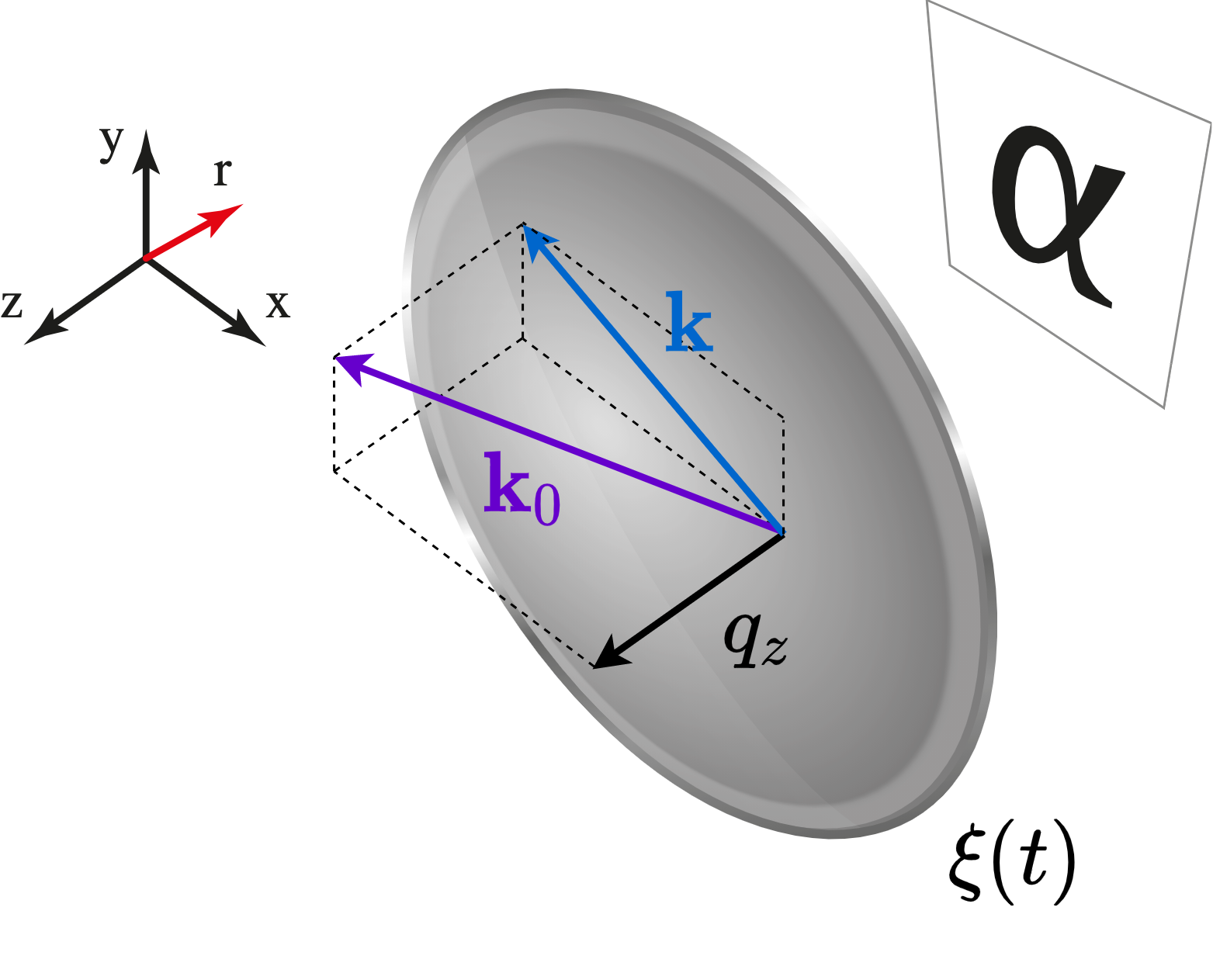}  
		\caption{}
		\label{fig832nma}
	\end{subfigure}
	\begin{subfigure}{.5\textwidth}
		\centering
		% include second image
		\includegraphics[width=.98\linewidth]{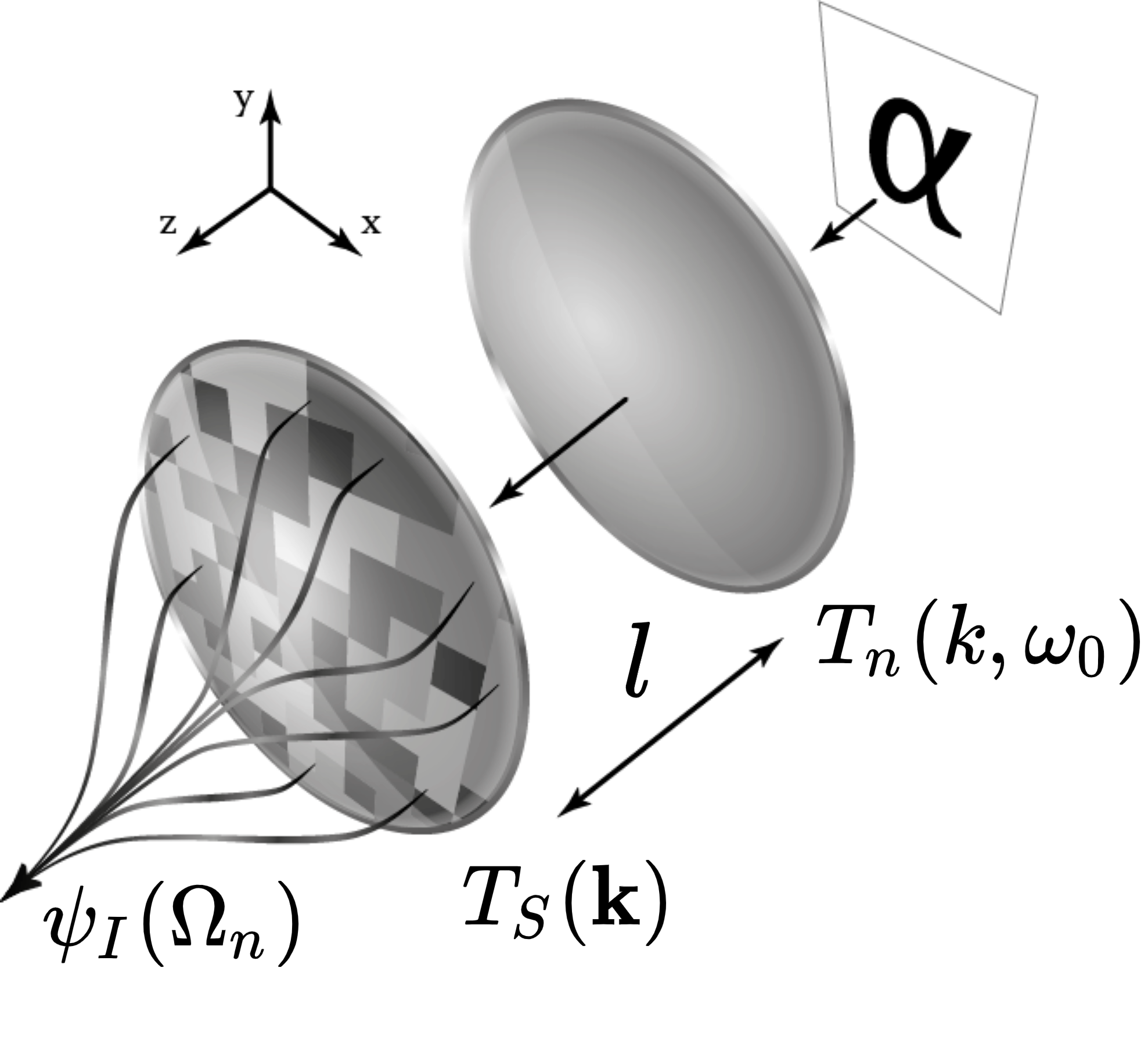}  
		\caption{}
		\label{fig832nmb}
	\end{subfigure}
	\caption{(a) Setup for imaging with a homogeneous time-modulated grating. (b) Setup for single-pixel imaging with a time-modulated grating combined with a second grating which can be identical to the first one or a random time-invariant receiving screen, depending on the application.}
	\label{fig832nm}
\end{figure}

For the recovery of the image, a ``mirror-grating'' configuration is used. The fundamental idea is illustrated in Fig. \ref{fig832nmb}, where an identical (i.e. $T_S (\fK) = T_n (k, \omega_0)$) mirror grating placed at a distance $l$ from the object and the imaging grating, is used as a recovery system, since it will couple all the information of the object into the fundamental component of the field. Then, after the mirror grating we will have
\beq
\psi_I(\fr,t)=\sum_{m=-\infty}^{\infty}e^{-i\Omega_mt}\iint B_m(\fK)e^{i\fK\cdot\fr}d^2\fK
\eeq
with
\beq
\label{eq:Bm}
B_m(\fK)=A_0(\fK)\sum_{n=-\infty}^{\infty}T_n(k)T_{m-n}(k) e^{i q_n l},
\eeq
where obviously only the terms for which $q_n$ is real will be relevant for the sum. Thus, if we measure the spatial field distribution corresponding to the source frequency $\omega_0$ and perform its spatial Fourier transform we obtain the $B_0(\fK)$ coefficients from which we can retrieve the source's Fourier transform from the above equation for $m=0$.

%\begin{figure}[h!]
%	\includegraphics[width=0.5\linewidth]{figures/mirror_grating}
%	\caption{...}
%	\label{fig2}
%\end{figure}
%

%\begin{figure}[h!]
%	\includegraphics[width=0.98\linewidth]{figures/results_var_pulse_width_1_20_fig}
%	\caption{to be adjusted}
%	\label{}
%\end{figure}

%\begin{figure}[ht]
%	\begin{subfigure}{.5\textwidth}
%		\centering
%		% include first image
%		\includegraphics[width=.98\linewidth]{figures/results_var_pulse_width_1_20_fig}  
%		\caption{}
%		\label{fig55a}
%	\end{subfigure}
%	\begin{subfigure}{.5\textwidth}
%		\centering
%		% include second image
%		\includegraphics[width=.98\linewidth]{figures/modulation_intensity}  
%		\caption{}
%		\label{fig55b}
%	\end{subfigure}
%	\begin{subfigure}{.5\textwidth}
%	\centering
%	% include second image
%	\includegraphics[width=.98\linewidth]{figures/random_modulation}  
%	\caption{}
%	\label{fig55c}
%\end{subfigure}
%	\caption{(a) $B_0 (|\fk|) / A (\fk)$ for modulation intensity $\mu = 10$, for various modulating pulse widths ($\nu_M = 0.5 \omega_0 / (2 \pi)$). (b) normalized maximum wavenumbers transferred through the grating system for pulse widths $T_m = \{ 1, \frac{1}{4}, \frac{1}{16} \}$ and for various modulation intensities $\mu$. \pp{Figure (c) shows results for random modulation - we can either combine it wit the previous figure, or just use random modulation since it looks much better than others.}}
%	\label{fig55}
%\end{figure}

The condition to fully recover the image is that we are able to excite a large enough number of temporal harmonics to propagate maximum wavenumber $k_\text{max}$, but as long as this happens we will have all the information in the fundamental mode at the mirror grating, without being required the  spatial analysis of higher harmonics. It is important to note that with this configuration we only need to excite higher harmonics at the first grating, but not to detect them at the mirror screen, so that perfect imaging can be done by the analysis of the field at the original frequency $\omega_0$ only. 

\begin{figure}[ht]
	\begin{subfigure}{.5\textwidth}
		\centering
		% include first image
		\includegraphics[width=.98\linewidth]{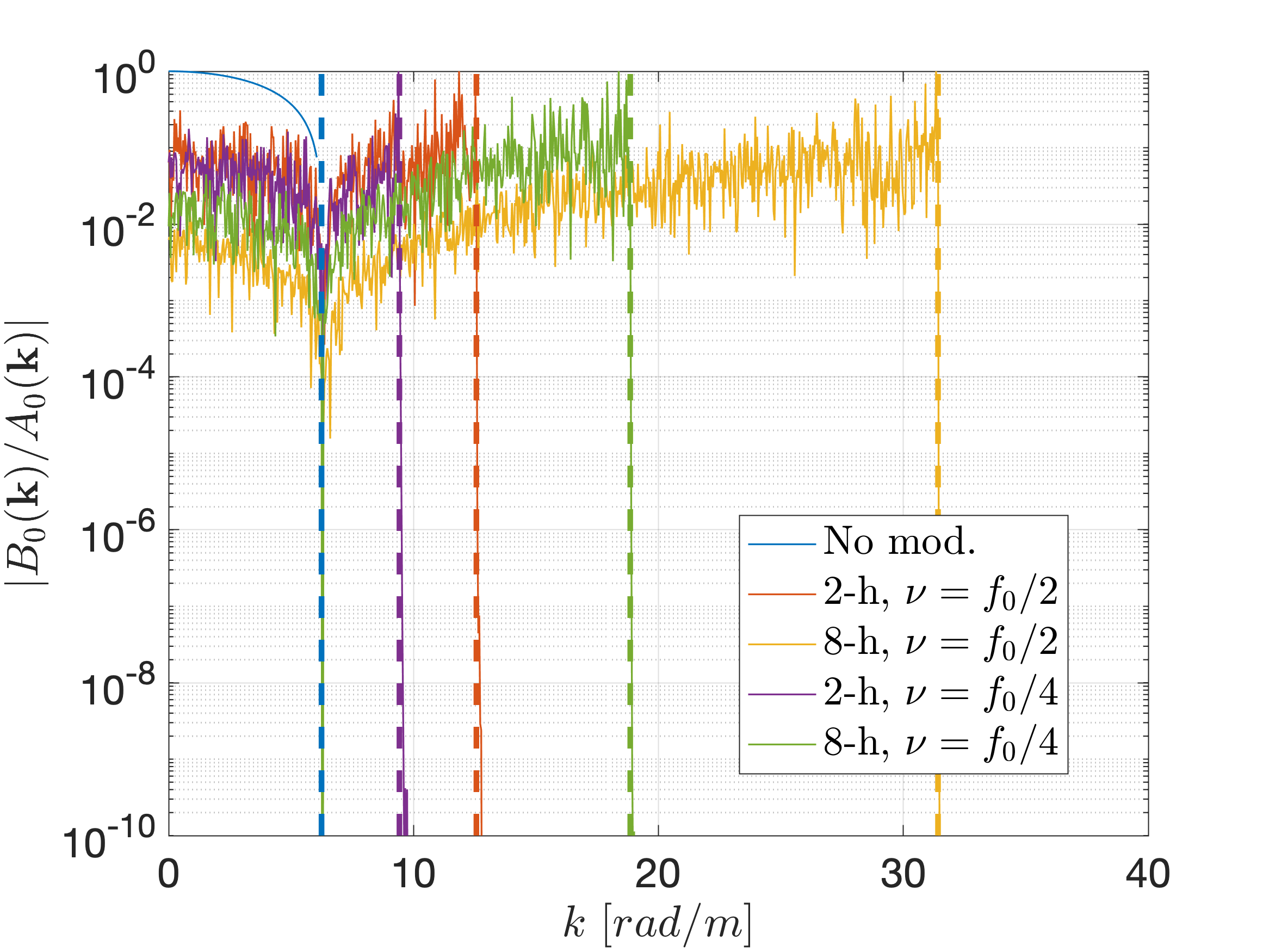}  
		\caption{}
		\label{fig55a}
	\end{subfigure}
	\begin{subfigure}{.5\textwidth}
		\centering
		% include second image
		\includegraphics[width=.98\linewidth]{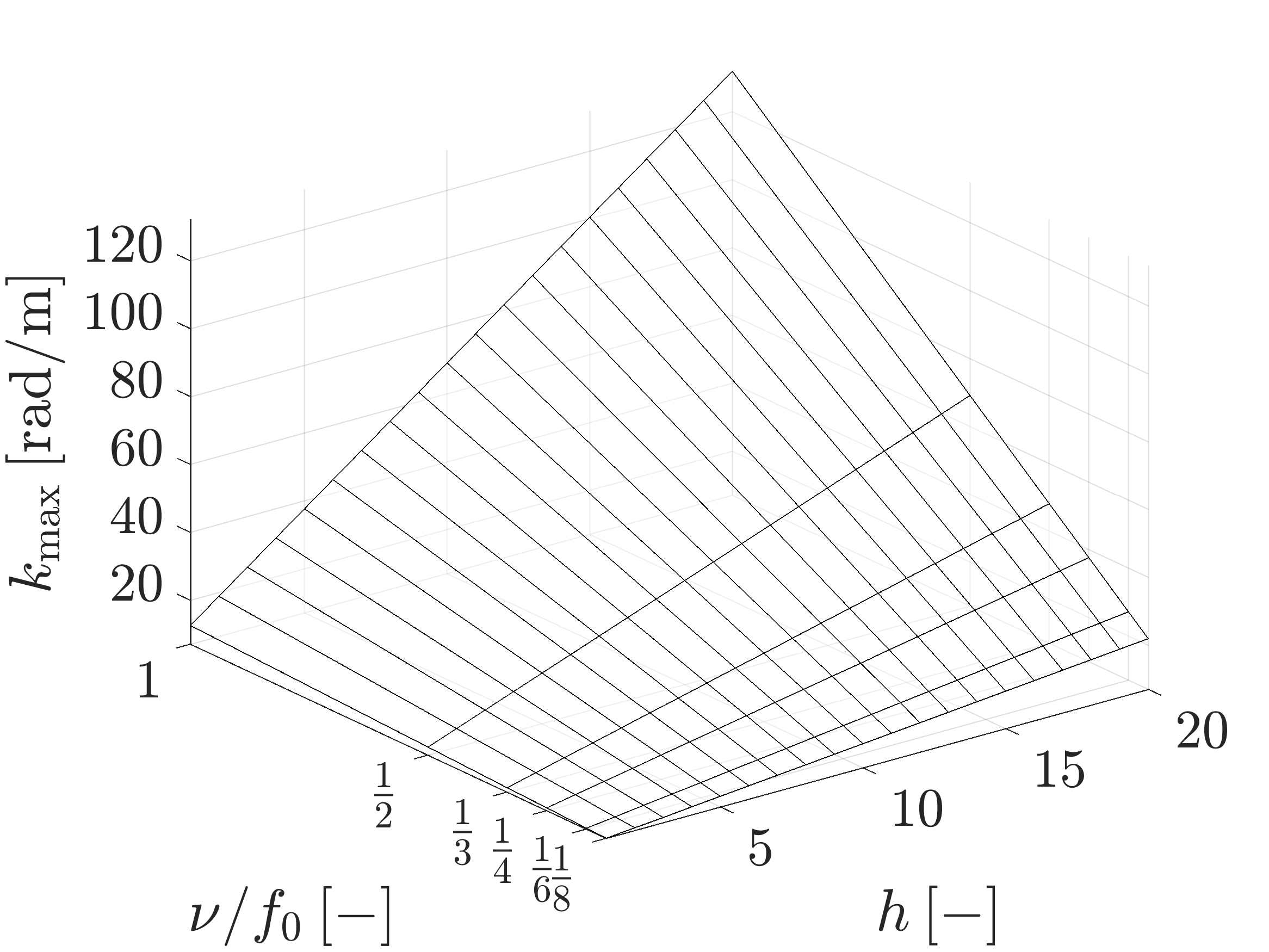}  
		\caption{}
		\label{fig55b}
	\end{subfigure}
	\caption{(a) The ratio $|B_0 (\fk) / A_0 (\fk)|$ (see Eq. \eqref{eq:Bm}) for the field frequency $\omega_0 = 1$ rad/s and  modulation frequencies $f_0 / 2$ and $f_0 / 4$ and for 2 and 8 excited higher harmonics. Vertical lines show corresponding cut-off wavenumbers $k_\text{max}$. (b) Maximum wavenumbers $k_\text{max}$ transferred through the grating system as a function of number of excited harmonics $h$ and modulation frequency $\nu$ ($f_0 = \omega_0 / (2 \pi)$).}
	\label{fig55}
\end{figure}

We demonstrate the application of the proposed method with the following example. Assuming unit field frequency, $\omega_0 = 1 \hspace{2pt} \text{rad/s}$, and $c_0 = 1$ m/s, we generate real-valued time-modulation signals with desired number of harmonics with their amplitudes and phases described by complex numbers randomly generated from the uniform distribution, $\xi_{|n|} \sim U (-1 - 1 i, 1 + 1 i)$. The maximum wavenumbers for waves transmitted through the imaging system for modulation signals that excite 2 and 8 harmonics, for two modulation frequencies ($f_0/2$ and $f_0/4$; $f_0 = \omega_0 / (2 \pi)$) are illustrated in Fig. \ref{fig55a}. It can be further seen from Fig. \ref{fig55b} that the maximum wavenumber to be transmitted is a bi-linear function of the modulation frequency and the number of harmonics excited by the modulation signal, $k_\text{max} (\nu, h)$. It is therefore preferred to use high modulation frequencies and/or detect possibly large number of harmonics in order to increase the quality of image reconstruction.

\begin{figure}[h!]
	\includegraphics[width=0.98\linewidth]{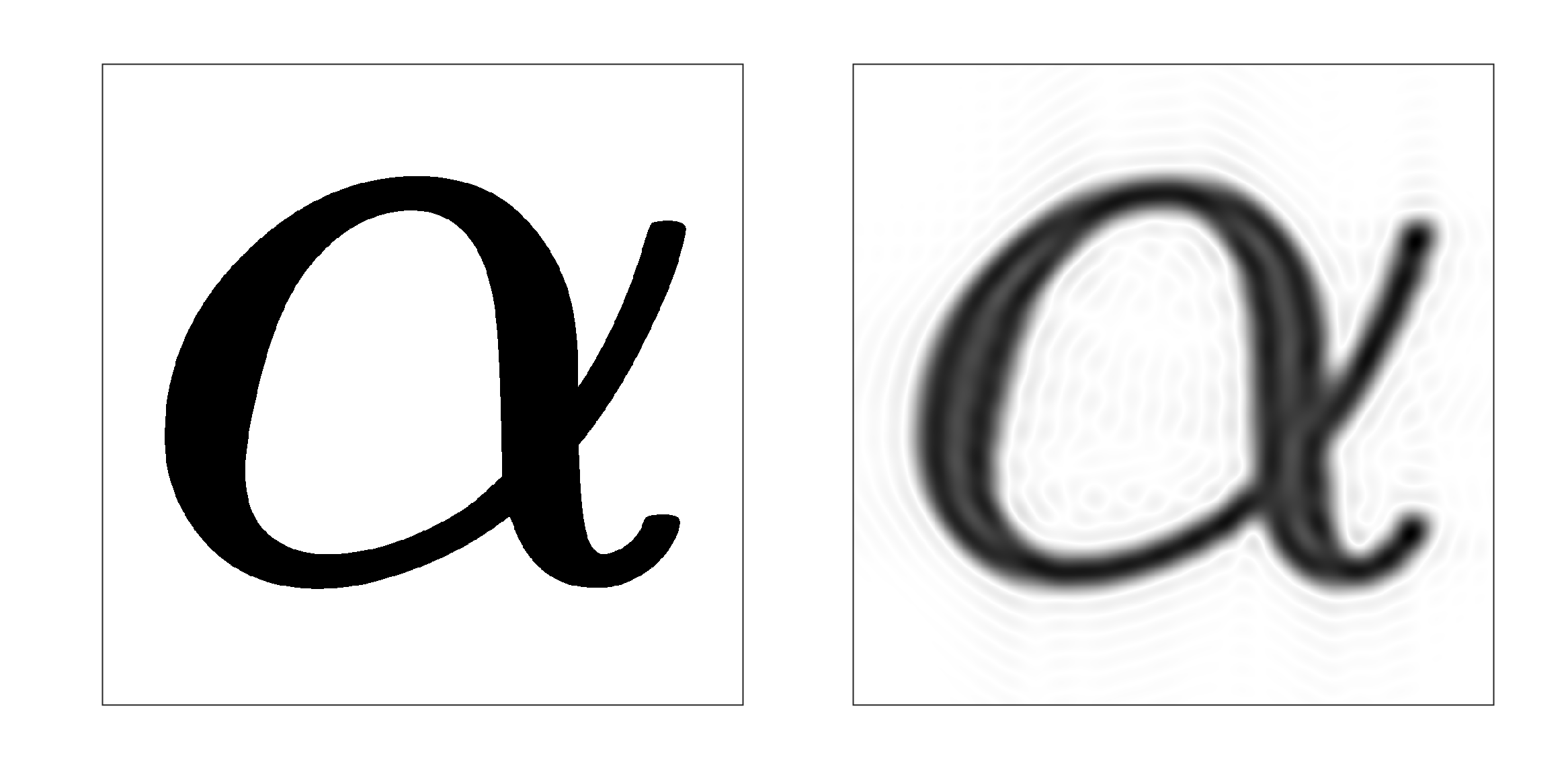}
	\caption{The model letter $\alpha$ used in numerical experiments (a), and its representation with $(1+128) \times (1+128)$ samples in the Fourier spectrum (b).}
	\label{fig7772}
\end{figure}

\begin{figure}[h!]
	\includegraphics[width=0.98\linewidth]{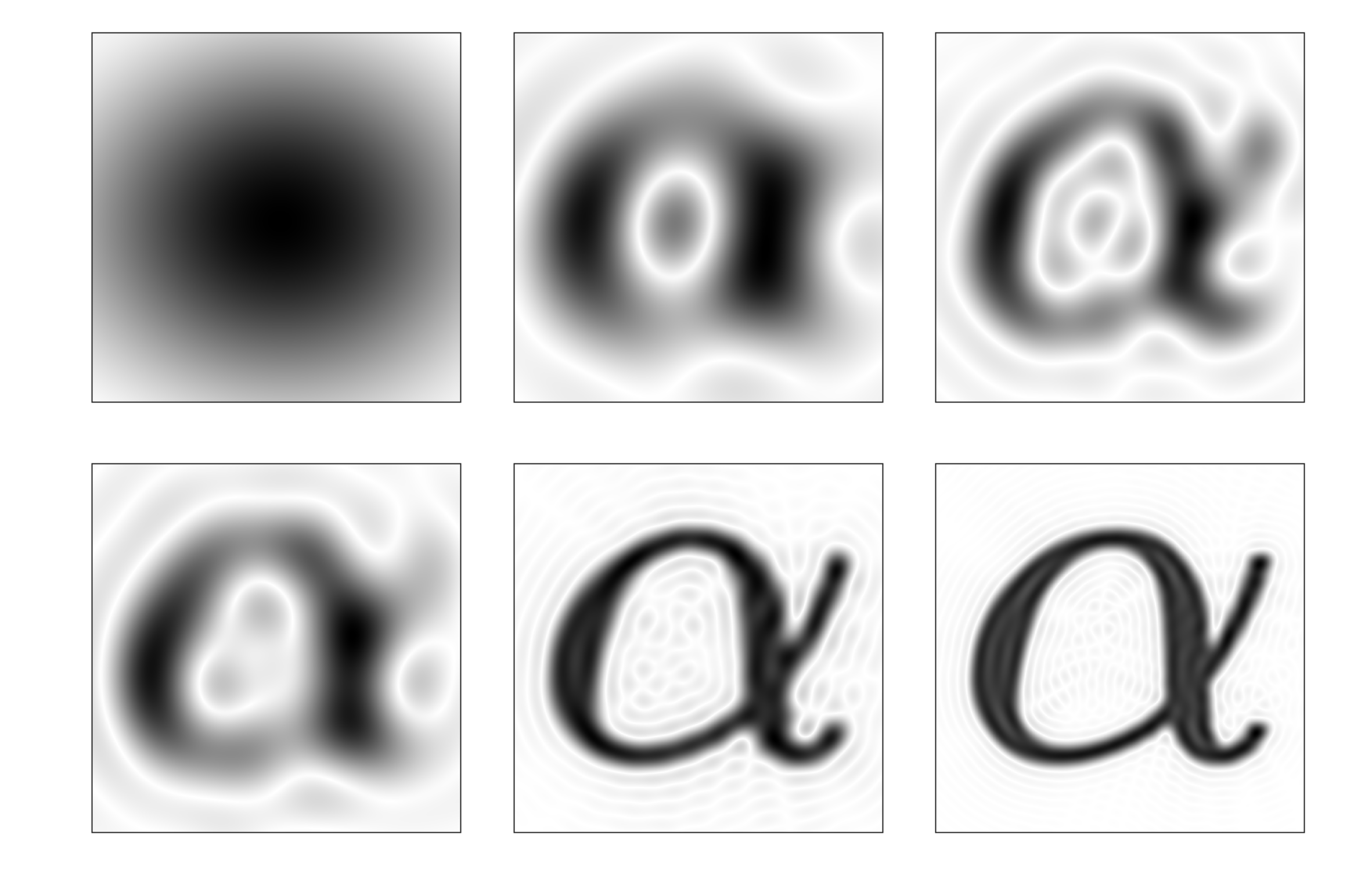}
	\caption{Recovered images of the letter (from left to right) obtained for $k_\text{max}$ for no modulation (left column) and 8 harmonics with $\nu = \{ f_0 / 4, f_0 / 2 \}$ (middle and right columns) for image size $\lambda \times \lambda$ (top row) and $4 \lambda \times 4 \lambda$ (bottom row). }
	\label{fig74}
\end{figure}

In order to illustrate how the maximum wavenumber, $k_\text{max}$, enhanced by the imaging with time-modulated gratings, influences image reconstruction we show in Fig. \ref{fig7772} an example letter that will be processed through the proposed system. Assuming that the object is illuminated by a field of unit frequency $\omega_0 = 1$ rad/s propagating with $c_0 = 1$ m/s, the letter is spatially Fourier-transformed and its spectrum truncated to contain wavevectors only up to $k_\text{max}$ that corresponds to a given modulation signal (8 harmonics for $f_0/4$ and $f_0/2$ modulation frequencies, corresponding to Fig. \ref{fig55a}). The reconstruction process consists of inverse Fourier-transforming of the truncated spectrum. The results are presented in Fig. \ref{fig74} for the same letters $\alpha$ assuming their physical dimensions of $\lambda \times \lambda$ (top row of Fig. \ref{fig74}) and $4 \lambda \times 4 \lambda$ (bottom row of Fig. \ref{fig74}), $\lambda = c_0 / f_0$. It is clear that for a letter size equal to the wavelength and no modulation, the reconstructed image does not reveal any features of the original pattern. For only 8 harmonics in the signal, however, and relatively low modulation frequencies, the shape of $\alpha$ can be clearly distinguished, especially for $\nu = f_0/2$. When the wavelength is four-times smaller (bottom row of Fig. \ref{fig74}), the letter is distinguishable for no modulation and almost perfectly reconstructed with 8 harmonics and modulation frequency of $\nu = f_0/2$.

%{\color{red} Plot of B0/A0 for periodic modulation and gaussian modulation?  It is a 1D plot, so we can put something}

%{\color{red} Image of a letter, its A0(k) and B0(k) and psiI 2x2 panel?}

%%%%%%%%%%%%%%%%%%%%%%%%%%%%%%%%%%%%%%%%%%%%%%%%%%%%%%%%%%%%%%%%%%%%%%%%%%%%%
%%%%%%%%%%%%%%%%%%%%%%%%%%%%%%%%%%%%%%%%%%%%%%%%%%%%%%%%%%%%%%%%%%%%%%%%%%%%%
%%%%%%%%%%%%%%%%%%%%%%%%%%%%%%%%%%%%%%%%%%%%%%%%%%%%%%%%%%%%%%%%%%%%%%%%%%%%%
\section{Single-detector imaging with time gratings}
%%%%%%%%%%%%%%%%%%%%%%%%%%%%%%%%%%%%%%%%%%%%%%%%%%%%%%%%%%%%%%%%%%%%%%%%%%%%%
The time grating allows an additional way for recovering the image without performing such a spatial scan. The idea consists in using all the other harmonics $\Omega_n$ to obtain spatial information from the object without scanning the sample.  The transversally homogeneous (the impedance uniformly distributed in space) nature of the time grating does not allow to distinguish between different orientations of the image, but we can replace the second grating by an in-homogeneous receiving screen of transmittance $T_S(\fr)$. The total field after this screen can be collected and integrated, so that a single time-dependent signal is received, as illustrated in Fig. \ref{fig832nmb}. If we measure the spectrum of this integrated signal, we will observe peaks at the different harmonics $\Omega_n$. Then the field distribution of the $n$-th harmonic after this screen will be
\beq
\psi_I(\Omega_n)=\iint d^2\fr T_S(\fr)\iint A_0(\fK)T_{n}(k,\omega_0)e^{i\fK\cdot\fr}e^{iq_nl}d^2\fK ,
\eeq
%%%%%%%%%%%%%%%%%%%%%%%%%%%%%%%%%%%%%%%%%%%%%%%%%%%%%%%%%%%%%%%%%%%%%%%%%%%%%\º
which is equivalent to
\beq
\psi_I(\Omega_n)=\iint T_S^*(\fK) T_{n}(k,\omega_0)e^{iq_nl} A_0(\fK)d^2\fK ,
\eeq
where $T_S(\fK)$ is the Fourier transform of the transmittance of the screen. The above equation can be seen as a linear system of equations of the form
\beq \label{nsha}
y_n=\sum_mM_{nm}A_m ,
\eeq
with the quantities $y_n$ being the measurement points $\psi_I(\Omega_n)$ (amplitudes and phases of harmonics) and $A_m$ are the unknown coefficients $A_0(\fK)$, which have been discretised in a set of $\fK_m$ elements and ordered in a vector labeled by $m$. Similarly, the matrix $M_{nm}$ is defined as
\beq \label{83ns}
M_{nm}=T_{n}(k_m,\omega_0) T_S^*(\fK_m) e^{i q_n l} ,
\eeq
and the knowledge of this matrix will allow us to recover all the $A_m$ from the spectrum of a signal received in a single detector. This approach is similar to single-pixel imaging methods, with the notable difference that all the information is encoded in different harmonics. Preferably, a possibly large number of harmonics is required in order to fully recover the components of $A_0(\fK_m)$. It is known, however, that images are sparse in the $\fK$ space, hence large $k_\text{max}$ does not in general mean that large $n$ in Eq. \eqref{nsha} is necessary. Also, it has to be pointed out that the screen $S$ should not have any inversion symmetry, since this would imply the column vectors of $M_{nm}$ corresponding to $\fK$ and $-\fK$ are identical and, consequently, the matrix $M_{nm}$ would not be invertible. The latter is indeed why the second receiving screen is required to recover the image from one single measurement. Note that since the time-modulated grating is homogeneous in space, only axially symmetric images could be recovered. Also, since the time grating is modulated externally, in the case of not having enough harmonics to recover the image, the modulation pattern (coding sequence) can be changed and the size of both the measurement set $y_n$ and the matrix $M_{nm}$ may be increased, since the $T_n$ elements will depend on the shape of the modulation. 

Finally, it has to be pointed out that the exponential term in Eq. \eqref{83ns} may vanish for $q_n = \sqrt{k_n^2 - k^2}$, $k_n = \sqrt{\Omega_n^2 / c_0^2} < k$, i.e. for small $\Omega_n = \omega_0 + 2 \pi n \nu$. Then, clearly, for a modulating frequency being an $r^{\text{th}}$ integer multiple of $\omega_0$, we have $\Omega_{n} = 0$ for $n = -r$ and, consequently, $q_n = i k$ and $e^{i q_n l} = e^{- k l}$. For practical setups, the latter exponential vanishes, lowering the rank of $M_{nm}$. It may be also concluded, that higher and incommensurate (with $\omega_0$) modulation frequencies are favorable for the coding sequences.

Below we present two approaches for recovering the image from the proposed single-pixel imaging system: the direct image recovery by directly solving the system \eqref{nsha} and a compressive sensing-based recovery using optimization. In both cases we use an inhomogeneous receiving screen with real-only impedance uniformly randomly distributed in space. The coding sequences are real-valued time-modulation signals of \eqref{nsy}, where 16 harmonics ($n = \{ 0, 1, ... 15 \}$) are used and their amplitudes and phases are described by complex numbers randomly generated from the uniform distribution, $\xi_{|n|} \sim U (-1 - 1 i, 1 + 1 i)$.

We denote a single measurement a procedure consisting of sending an input modulating sequence through the system and acquiring the temporal Fourier spectrum of the response composed of all signals integrated after passing the receiving screen. From this spectrum, the total number of 31 harmonics (amplitudes and phases; $n$ rows of $M_{nm}$ matrix) are acquired and stored as a column of $M_{nm}$. Then, the procedure is repeated for another coding sequence and another column is appended to $M_{nm}$. 

In numerical experiments reported in this section, we consider an image of a letter illustrated in the left panel of Fig. \ref{fig7772}, of size $1024 \times 1024$ pixels. Physical dimension of the letter panel is $\lambda \times \lambda$. For recovery, we use a limited part of the 2-D Fourier spectrum of the letter consisting in $(1 + 128) \times (1 + 128)$ samples (corresponding to the DC and $\pm 64$ samples in each direction). The letter reconstructed by using all $129 \times 129$ samples in the truncated spectrum is presented in the right panel of Fig. \ref{fig7772}. Due to the symmetry of the spectrum, the letter is uniquely defined by 8385 complex coefficients $A_m$.

%\subsection{Some properties of $M_{nm}$}
%
%For compressive sensing it is preferable that $M_{mn}$ is a random matrix. Each column of the matrix (i.e. corresponding to its $m^\text{th}$ index) is related to a single value of the wavevector $k_m$. It can be noted that the $m$ indices corresponding to $k_m = \{ -2 \pi , 0, 2 \pi \}$ (for $\omega_0 = 2 \pi$ and the side length of the letter panel equals $\lambda_0$).

\subsection{Direct image recovery}

For the direct recovery of the image we require the matrix $M_{nm}$ to be square and invertible to recover $A_0$ through $A_m$ coefficients. Consequently, a possibly large number of harmonics and substantial number of measurements will be necessary. We consider 30 harmonics, $n = \{ 1, ..., 15 \}$, and take 560 measurements obtaining square $M_{nm}$ and the measurement vector $y_n$. Through the standard Gauss elimination process we compute $A_m$ from \eqref{nsha}. The resultant recovered image of the letter is shown in the bottom right end panel of Fig. \ref{fig9937}. The reconstruction process is lossless for the direct recovery.

%\begin{figure}[h!]
%	\includegraphics[width=0.98\linewidth]{figures/single_pixel_letters_cs}
%	\caption{SPACE HOLDER - this figure will show direct reconstruction.}
%	\label{fig5243}
%\end{figure}

\subsection{Compressive sensing of the image}

\begin{figure}[h!]
	\includegraphics[width=0.98\linewidth]{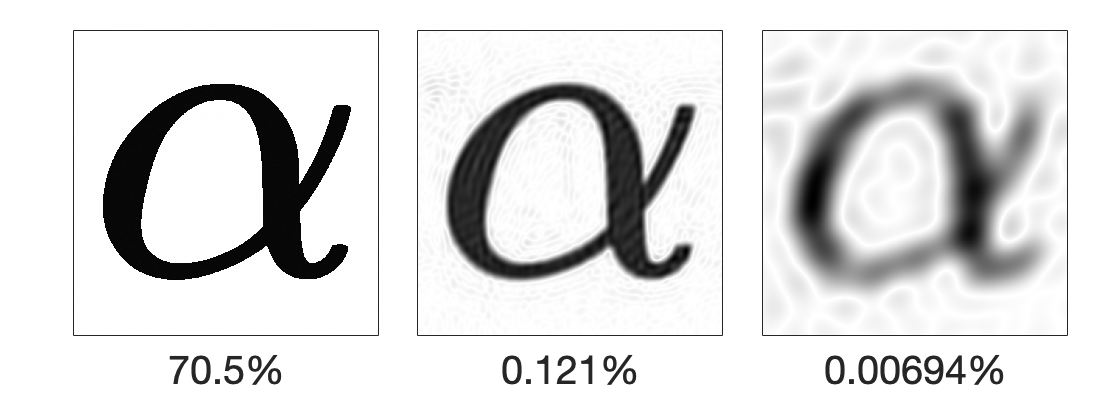}
	\caption{The letter $\alpha$ reconstructed from a limited number of samples (given below the plots in $\%$ up to three significant digits) in the $\fk$ space.}
	\label{fig83b9}
\end{figure}

The direct recovery of the image requires a number of measurements to be taken before the problem in \eqref{nsha} can be solved. Although, the process is relatively simple and fast, the number of measurements can be substantially reduced due to the sparsity of the image in the wavenumber domain. The sparsity of the example letter $\alpha$ used in the reconstruction can be clearly verified by inspecting Fig. \ref{fig83b9}, where the letter was reconstructed only from a very limited number of the most significant samples (highest $|A_m|$) in the $\fk$ space.

For signals having sparse representation in some space (the Fourier space in this case), as $A_m$, the compressive sensing theory \cite{Baraniuk2007,Thiagarajan2014} can be used to solve Eq. \eqref{nsha} for substantially smaller number of measurements than for the direct reconstruction, i.e. for $M_{nm}$ being a rectangular matrix with $n \ll m$. In the compressive sensing process, only the most significant sparse coefficients of the representation are recovered.

%\beq \label{xuen}
%\hat{A}_m = argmin || A'_m ||_1 \hspace{6pt} \text{such that} \hspace{6pt} y_n = M_{mn} A'_m ,
%\eeq

The properties of the measurement matrix $M_{nm}$ are fundamental to proper reconstruction \cite{Baraniuk2007}. Therefore, we construct $M_{nm}$ from repeated random measurements and aim at satisfying the restricted isometry property and incoherence between measurements. The solution of \eqref{nsha} for $n \ll m$ is found through the matching pursuit algorithm \cite{Mallat1993}, by requiring
\beq \label{xuen}
\text{min} \hspace{2pt} || y_n - \sum_m M_{nm} \hat{A}_m || ,
\eeq
where $\hat{A}_m$ is found iteratively by computing the correlation vector at $j$-th iteration as
\beq \label{xuenua}
c^j_o = \sum_n M_{no}^* (y_n - \sum_m M_{nm} \hat{A}^j_m),
\eeq
and updating the current approximation to the sparse representation $\hat{A}^{j+1}_m = \hat{A}^{j+1}_m + c_s \delta_s$, where the index $s$ maximizes the correlation, i.e. $\text{max}_s \hspace{2pt} |c_s|$. As a result, each iteration localizes and approximates a sparse coefficient in $\hat{A}_n$.

\begin{figure}[h!]
	\includegraphics[width=0.98\linewidth]{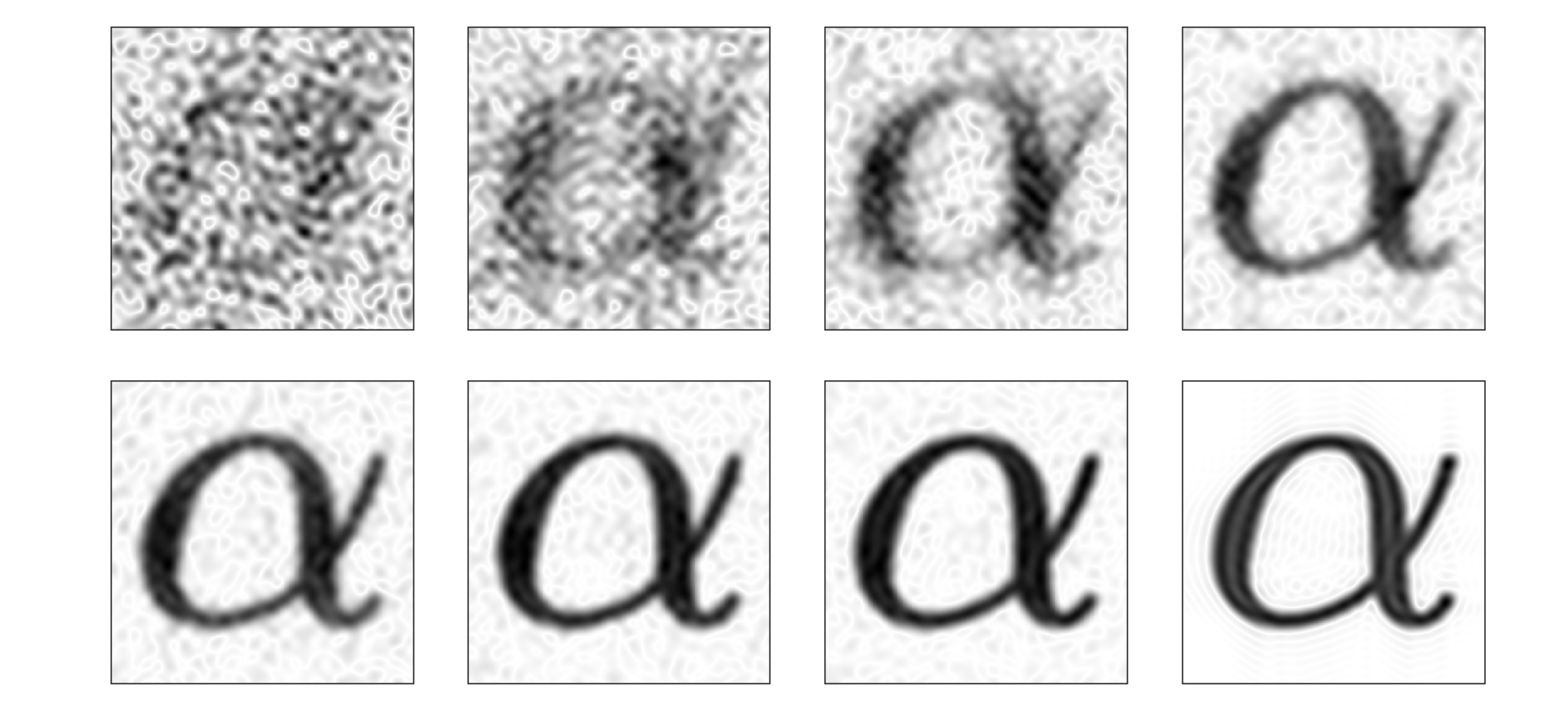}
%	\caption{The letter $\alpha$ composed of $m = 8880$ discrete wavevectors $\textbf{k}_m$ reconstructed from 10 measurements ($n = 410$, left) and 20 measurements ($n = 820$, right).}
\caption{Letters reconstructed by the compressive sensing approach by using (from top left): 25, 50, 100, 200, 300, 400 and 500 experiments. The last plot (lower right) shows the result of direct reconstruction (corresponding to 560 experiments).}
	\label{fig9937}
\end{figure}

The results of the single-pixel reconstruction process for 25, 50, 100, 200, 300, 400 and 500 measurements with 15 harmonics are presented in Fig. \ref{fig9937}. Compared to the direct image reconstruction that required 560 measurements, it can be seen that using only 100 experiments (i.e. less than 20$\%$ of data), the letter can be clearly recognized.

The possibility of recovering the image of an object from a single-point measurement is more than relevant for those fields where large arrays of detectors are especially expensive, as is the case of microwaves or acoustic waves, where the low frequency of the fields makes it more realistic to physically implement a time-modulated material. Although great advances are currently done in the  temporal modulation of materials in the optical domain, the excitation-detection of a large enough number of harmonics to be able to perform perfect imaging for optical waves is something extremely challenging. Therefore, we consider that a realistic proposal for a realization of this imaging system should begin for low frequency fields.

%%%%%%%%%%%%%%%%%%%%%%%%%%%%%%%%%%%%%%%%%%%%%%%%%%%%%%%%%%%%%%%%%%%%%%%%%%%%%
%%%%%%%%%%%%%%%%%%%%%%%%%%%%%%%%%%%%%%%%%%%%%%%%%%%%%%%%%%%%%%%%%%%%%%%%%%%%%\º
\section{Summary}
%%%%%%%%%%%%%%%%%%%%%%%%%%%%%%%%%%%%%%%%%%%%%%%%%%%%%%%%%%%%%%%%%%%%%%%%%%%%%
In summary, we have shown that a time-modulated grating can be used to overcome the diffraction limit for imaging in the far field. It has been shown that placing two identical gratings between the object and the detector produces a field distribution from which we can recover the image of the object with virtually unlimited resolution, since the evanescent components of the image have been coupled to high-frequency harmonics by the first grating and decoupled by the second one. If the object is axially symmetric, this method allows for the recovery of the image by analyzing the temporal spectra of the total field at a single point, being therefore a super-resolution single-pixel imaging system. In the case of having non-symmetric objects, the second time-grating can be replaced by a spatially inhomogeneous screen, and the single-pixel recovery method works similarly. Finally, the theory of compressive sensing has been used to reduce the number of required data significantly, and numerical experiments have been presented supporting our findings. This work shows that space-time gratings can work as ultra-fast single-pixel perfect imaging devices, which opens the door to a new set of applications not only in optics but also in other domains using waves for imaging.

%%%%%%%%%%%%%%%%%%%%%%%%%%%%%%%%%%%%%%%%%%%%%%%%%%%%%%%%%%%%%%%%%%%%%%%%%%%%%%%%%%%%%%%%%%%
%%%%%%%%%%%%%%%%%%%%%%%%%%%%%%%%%%%%%%%%%%%%%%%%%%%%%%%%%%%%%%%%%%%%%%%%%%%%%%%%%%%%%%%%%%%
\acknowledgments
Daniel Torrent acknowledges financial support through the ``Ram\'on y Cajal'' fellowship under grant number RYC-2016-21188 and to the Ministry of Science, Innovation and Universities through Project No. RTI2018- 093921-A-C42. Pawel Packo acknowledges the support of National Science Centre in Poland through grant no. [2018/31/B/ST8/00753]. This research was supported in part by PLGrid Infrastructure. Both authors acknowledge Steven Cummer, Jes\'us Lancis, Enrique Tajahuerce and Vicente Dur\'an for useful and fruitful discussions. 

%\bibliography{bibliography}
%merlin.mbs apsrev4-1.bst 2010-07-25 4.21a (PWD, AO, DPC) hacked
%Control: key (0)
%Control: author (8) initials jnrlst
%Control: editor formatted (1) identically to author
%Control: production of article title (-1) disabled
%Control: page (0) single
%Control: year (1) truncated
%Control: production of eprint (0) enabled
%

\end{document}